\documentclass{rsauthor}
\usepackage{amssymb,epsfig,setspace,latexsym,graphicx,rotating,textcomp,amsmath,wrapfig,multirow,booktabs,tabularx}
\usepackage[numbers,sort&compress]{natbib}
\usepackage[hang,small,bf]{caption}
\usepackage[a4paper,left=0cm,right=5cm]{geometry}

\begin{document}

\title{Density-functional theory study of gramicidin A ion channel geometry and electronic properties}
\author{Milica Todorovi\'c$^{1}$, D. R. Bowler$^{2,3,4}$, M. J. Gillan$^{3,4}$, Tsuyoshi Miyazaki$^{1}$}
\address{$^{1}$National Institute for Materials Science, 1-2-1 Sengen,
  Tsukuba, Ibaraki 305-0047, Japan\\
$^{2}$International Centre for Materials Nanoarchitectonics (MANA), National
Institute for Materials Science, 1-1 Namiki, 
  Tsukuba, Ibaraki 305-0044, Japan\\
$^{3}$London Centre for Nanotechnology,
UCL, 17-19 Gordon Street, London WC1H 0AH, UK\\
$^{4}$Thomas Young Centre and Department of Physics \& Astronomy, UCL, Gower Street, London WC1E 6BT, UK}
\date{}
\abstract{Understanding the mechanisms underlying ion channel function from the atomic-scale requires accurate ab initio modelling as well as careful experiments. Here, we present a density functional theory (DFT) study of the ion channel gramicidin A, whose inner pore conducts only monovalent cations and whose conductance has been shown to depend on the side chains of the amino acids in the channel. We investigate the ground-state geometry and electronic properties of the channel in vacuum, focusing on their dependence on the side chains of the amino acids. We find that the side chains affect the ground state geometry, while the electrostatic potential of the pore is independent of the side chains.  This study is also in preparation for a full, linear scaling DFT study of gramicidin A in a lipid bilayer with surrounding water. We demonstrate that linear scaling DFT methods can accurately model the system with reasonable computational cost. Linear scaling DFT allows ab initio calculations with 10,000 to 100,000 atoms and beyond, and will be an important new tool for biomolecular simulations.} 
\keywords{ion channel, gramicidin, DFT, linear scaling, electronic structure calculation}
\maketitle

\section{Introduction}
\label{sec:introduction}
\begin{figure}
\begin{center}
\includegraphics[scale=0.1,clip=]{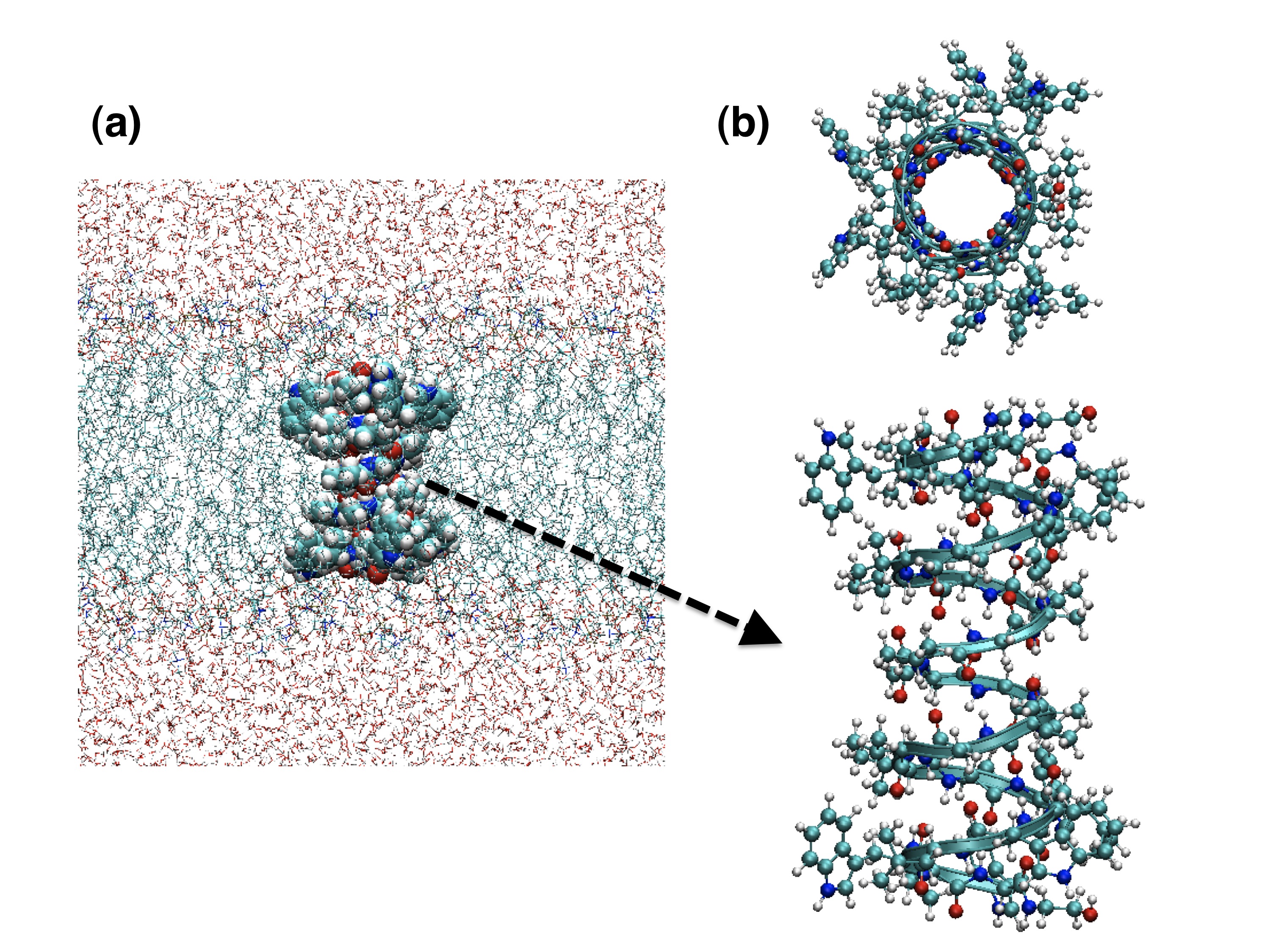} 
\caption{(a)Structure of gramicidin-A ion channel in lipid bilayers. The upper and lower
area are bulk water region. (b) Top (upper) and side (lower) views of the gramicidin-A molecule.
(Online version in colour.) }
\label{SchematicFigure}
\end{center}
\end{figure}

Ion channels reside in cell membranes and control the passage of ions, a vital function in regulating cell behaviour.
The gramicidin A (gA) ion channel is  one of the smallest and simplest ion channel systems\cite{AndersenBook,Maffeo:2012CR6250,Kelkar:2007aa}, 
and was the target of the first antibiotic used in clinical practice.  
It is formed from a stack of two short beta-helix polypeptides, 
as shown in Fig.~\ref{SchematicFigure}. Each polypeptide contains fifteen amino acids in a sequence of 
alternating left-handed and right-handed types. 
The latter helicity is rarely found in natural peptides, and owing to the alternating 
helicity, all peptide residues face outward into the lipid region, forming the unique 
helix structure with the beta type backbone structure.
As is seen in Fig.~\ref{SchematicFigure}, the channel structure forms a narrow pore, through which 
water molecules or ions can permeate in single file. 

In spite of its simple structure, gA shows high selectivity towards ion permeation.
It is selective for monovalent cations, such as H$^{+}$, K$^{+}$, Na$^{+}$, or Cs$^{+}$, showing no measurable permeability for anions or polyvalent cations\cite{Myers1972313}. 
Among many ion channels, gA was one of the first whose atomic resolution structure was available. 
Many experiments have been performed to investigate its transport properties and 
to clarify the mechanism of its high selectivity. 
Considering the existence of significant experimental information as well as its simple structure, 
the gA system is of great importance as a model membrane protein\cite{Kelkar:2007aa}.

In the computational studies of the gramicidin-A, two high resolution structures have 
been commonly used so far. They are labelled as 1MAG and 1JNO in the PDB database, respectively. 
The former is obtained from solid-state NMR for the channel in DMPC membrane\cite{1997.ketchem.1655}, 
while the latter one is from solution NMR, in detergent micelles\cite{2001.townsley.11676}. 
While these two structures are topologically similar, they exhibit some variation 
in helix properties and residue Tryptophan 9 (Trp9) orientation (Fig.~\ref{mag_jno}(a)). 
The 1MAG structure helix has R=6.5 residues per helix turn and d=4.95~\AA\ helix turn 
height (pitch) in constrast to R=6.3 and d=4.79~\AA\ of the 1JNO structure. 
The structural disparities could have originated from any aspect of experimental procedure, 
with little to indicate the optimal channel geometry: a classical potential MD study 
comparing the two structures in a lipid membrane environment predicted that 
the two backbones after dynamic relaxation starting from two structures become 
almost the same and that Trp9 spends 80 \% of the time in the 1JNO orientation
and 20 \% in the 1MAG orientation\cite{2003.allen.9868}.   
Although we do not treat the ion permeation directly in this study, it is important to note 
the experimental result that the rate of the ion permeation is significantly changed 
by the replacement of the Tryptophan(Trp) residues, which have dipole moments ($\sim 2D$),
with other residues \cite{Becker:1991BC8830,Hu:1995BC14147,Andersen_ja980182l, 2001.townsley.11676},
When it is replaced with nonpolar Phenylalanine (Phe), the ion permeation rate is greatly reduced.
The change of the ion permeation rate is larger if the number of substituted Trp groups is larger, and also strongly depends 
on which Trp groups are replaced; the mutation of Trp9 is reported to show the largest change.

The system is an ideal target for theoretical studies. 
Although most theoretical studies were based on semi-microscopic models in the early stage,
it is necessary to employ atomic-scale simulations, such as molecular dynamics (MD),
for quantitative and detailed understanding of the permeation properties\cite{2002.edwards.1348}.
Many MD simulations of the gA ion channel system have been already 
reported\cite{1996.woolf.92,Woolf22111994, 1999.chiu.1939,2003.allen.9868,2003.allen.2159,Bastug:2006BJ2285} and 
nowadays it is common to treat the gA system in the channel environment, that is
gA with surrounding lipid bilayers (a simple model for the membrane) and bulk water.
Such MD studies have clarified many aspects of the system, but they have a serious problem
that the results sometimes depend on the type of the force field used in the calculations\cite{2003.allen.2159}. 
Since the gA system with its environment requires accurate modelling
of many different types of interaction,
the transferability of classical force fields may lead to poor
description of specific interactions.
For example, water molecules and ions are transported through the narrow pore of the gA channel.
It is unreliable to use an unpolarisable force field for water molecules or 
ions in such a unique constrained space\cite{Bucher:2009CPL207,Patel:2009aa}.
One of the well-known problems of theoretical studies using classical force fields is that 
the calculated energy barriers reported for the ion translocation 
in the gA pore are rather too high to explain the experimentally observed permeation rate\cite{Bastug:2006BJ3941,2003.allen.2159,2006.allen.3447}. 
It is therefore important to apply more accurate methods to the gA system.

\textit{Ab initio} approaches based on density functional theory (DFT) have been playing important 
roles in clarifying the physical properties of various kinds of
materials, though standard implementations of DFT are limited in the
system size which they can model. 
A simplified gA model system has been modelled by DFT calculations
to study proton diffusion in the pore of the gA channel\cite{Sagnella:1996fj}. 
A recent quantum mechanics / molecular mechanics (QM/MM) study 
on the gA system showed the importance of polarisation of the water molecules in the gA pore\cite{Bucher:2009CPL207, timko:205106}.
However, theoretical studies based on DFT are still very limited  
because of the demanding computational time and scaling with system
size.  Recent work has developed DFT methods which scale linearly with
system size\cite{Bowler:2012zt}, allowing DFT calculations on systems
with 10,000--100,000 atoms or more\cite{Bowler:2010uq}.  This brings the possibility to
perform DFT calculations on entire biological molecules and their environment.
In this work, we calculate the atomic and electronic structures of the
isolated gA using the linear scaling DFT code {\sc Conquest}\cite{2002.bowler.2781,2006.bowler.989}.
In particular, we investigate the effects of the side chains of the
channel on its structural and electronic properties.
Although we only treat the isolated gA here, we still obtain important information 
about the structural stability and electronic structure of the gA molecule itself.
We expect that such information will also help to 
clarify and improve the accuracy of force fields in the future.
To our knowledge, it is the first DFT study to treat the whole gA molecule 
without any simplification.  We will report linear scaling DFT
calculations on the gA channel with full environment in a future publication.

The rest of the paper is organized as follows.
In Sec.~\ref{sec:comp-meth-local}, we briefly explain the computational method used in this work. 
The results of our calculations on the atomic and electronic structures of the isolated gA 
molecule are presented in Sec.~\ref{sec:results-discussion}.  
Finally, concluding remarks are given in Sec.~\ref{sec:conclusions}.

\section{Computational Method: Local orbital and order-N methods}
\label{sec:comp-meth-local}
The calculations in this work are performed using the {\sc Conquest} code, which 
is designed for very large-scale DFT calculations\cite{2002.bowler.2781,2004.miyazaki.6186,2006.bowler.989,Bowler:2010uq}.
The method used in the code is based on density functional theory (DFT), with the pseudopotential 
technique. In this work, we employ the generalized gradient approximation (GGA) 
by Perdew, Burke and Ernzerhof (PBE)\cite{1996.perdew.3865}
for the exchange-correlation energy.

In {\sc Conquest} we work with local orbitals \{ $\phi_{i \alpha}(r)$ \}, which are 
centered on the position of each atom $i$ with the index of orbital $\alpha$.
We represent the Kohn-Sham orbitals $\Psi_{\nu} (r)$ ($\nu$: band index) as 
linear combinations of the local orbitals.
\begin{equation}
\Psi_{\nu} (r) = \sum_{i \alpha} c_{i \alpha}(\nu) \phi_{i \alpha} (r)
\end{equation}
These local orbitals, which we call support functions, are themselves expressed by 
basis set functions.  {\sc Conquest} provides two types of basis set functions: 
B-splines on regular grids for accurate calculations since the basis set is systematically
improvable\cite{1997.hernandez.13485}; and pseudo atomic orbital (PAO) basis set for efficient calculations\cite{2008.torralba.294206}.  
Users can choose one of the two basis set functions
depending on the purpose. We only use the latter in this study.

Although {\sc Conquest} can solve for the Kohn-Sham orbitals using a conventional $O(N^3)$ 
diagonalisation technique, the big advantage of using {\sc Conquest} is that it can also 
employ $O(N)$ calculations.
In the $O(N)$ calculations,  the code uses the density matrix minimization method proposed by 
Li, Nunes and Vanderbilt (LNV)\cite{McWeeny:1960zp,1993.li.10891}.
Since the detailed explanation of the method has already been given in our previous papers,
only the main points are explained in the following.
In the density matrix minimization method, we calculate the density
matrix instead of the Kohn-Sham orbitals.  Formally, the density matrix is defined as 
\begin{equation} 
\rho(r,r^{\prime})=\sum_{\nu} f_{\nu} \Psi_{\nu}^{\ast} (r) \Psi_{\nu} (r^{\prime}) .
\end{equation}
The total energy within DFT can be expressed by the density matrix $\rho$ and 
we optimise it instead of solving Kohn-Sham orbitals. 
In the LNV method, the matrix $K$ is expressed as $K=3LSL-2LSLSL$ 
to keep the density matrix weakly idempotent. 
Then, we introduce a cufoff radius $R_L$ for the off-diagonal elements of the $L$ matrix 
in order to utilize the locality of density matrix.
One important advantage of the LNV method is that it is variational with
respect to $R_L$. We can examine the cutoff energy dependence of total energy 
easily and can evaluate an error from using a finite cutoff of $R_L$. 
This convergence behavior is reported in 
Sec.~\ref{sec:results-discussion}(\ref{sec:order-n-calculations})
for the DFT calculation of the isolated gA system.

In both diagonalisation and $O(N)$ methods, the code can perform non-self-consistent (NSC)
DFT calculations using the Harris-Foulkes energy functional with a charge density constructed 
from the superposition of atomic charge\cite{Harris1985,Foulkes1989}. 
Although the NSC method is less accurate than usual DFT calculations with 
the self-consistent field (SCF) method, it enables us to perform efficient structure optimization
or molecular dynamics.  Note that the code can calculate forces which are completely consistent 
with the NSC method\cite{2004.miyazaki.6186,Torralba:2009JCTC1499}.
This method may be useful for our future study on gA systems, and its accuracy is also 
checked in this paper.

\section{Results and Discussion}
\label{sec:results-discussion}

\subsection{Structural Properties\label{sec:struct-prop}}
\begin{figure}
\begin{center}
\includegraphics[width=0.7\textwidth]{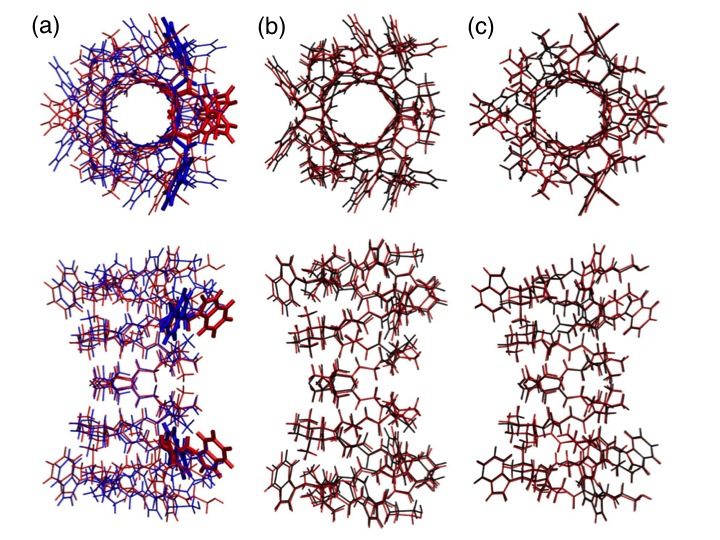} 
\caption{ (a) Comparison of 1MAG (blue/dark) and 1JNO (red/light) experimental structures. 
\textit{Ab initio} structural optimisation initial (black/dark) and final (red/light) structures in the (b) 1MAG case; (c) 1JNO case.
(Online version in colour.)}
\label{mag_jno}
\end{center}
\end{figure}

 In this section, we report our results on the structural properties of the gA molecule in the vacuum.
Here, we perform structure optimisation using two different high-resolution 
structures as initial atomic positions, 1MAG and 1JNO structures (Fig.~\ref{mag_jno}(a)). 
In order to employ efficient structure optimisation, we first use the NSC method, 
then switch to the usual SCF method\cite{Torralba:2009JCTC1499}. 
In both NSC and SCF methods, all atomic positions are relaxed using the standard conjugate 
gradient method until the maximum force component becomes smaller than 0.05 eV/~\AA.  
Note that the calculations in Sec.~\ref{sec:results-discussion}\ref{sec:struct-prop} and Sec.~\ref{sec:results-discussion}\ref{sec:electronic-structure} are done by the standard diagonalisation 
method with the DZP basis set, explained in Sec.~\ref{sec:comp-meth-local}, 
and GGA-PBE functional\cite{1996.perdew.3865}.
We use only a $\Gamma$ point for $k$-point sampling and the cutoff energy for the
charge density grid is 150 Hartree. Periodic boundary is used with a unit cell of
$ 23.3 $\AA $\times 23.3 $\AA $\times 35.0 $\AA. 

 Figure~\ref{mag_jno}(b) and (c) show the initial and optimised atomic positions starting from 1MAG 
and 1JNO structures, respectively. 
Although the original structures are obtained in different environment (in DMPC bilayers 
or in SDS micelles), we found that the optimised positions are close to the original ones in both cases.
These figures also show that 1MAG has larger differences between the initial and 
final coordinates, especially for the backbone structure.
 In order to investigate the backbone structure in detail, we calculate the number of residues 
per helix turn $R$ and the helix pitch $d$ following the method in Ref.\cite{Ketchem:1997STR1655}.
The root mean square deviations (RMSD) for the atoms of the backbone are also calculated.
These structural parameters of 1MAG and 1JNO, before and after the structure optimisation, 
are listed in Table~\ref{geometry}. 
The results show that the optimised 1MAG and 1JNO have different backbone structures.
In addition, the change of the backbone structure is small in 1JNO case, 
while it is significant in 1MAG case, especially for helix pitch $d$. 
Although the original value of $d$ in 1MAG case is already larger than that of 1JNO, 
the increase of $d$ in 1MAG case is 0.15~\AA, which is larger than 0.08~\AA\ in 1JNO case. 
We observe a similar behaviour in the hydrogen bond lengths. 
The $\beta$ -helix structure is stabilised by the hydrogen bonds formed between the backbone 
amide hydrogen and the carbonyl oxygen atoms parallel to the pore axis. 
The increase of the hydrogen bonds by the structure optimisation is 12\% in 1MAG, 
while it is 3\% in 1JNO case.

 
\begin{table}[t]
\centering
\begin{tabular}{cccccccccc}
\toprule
Structure (SC) & RMSD & & $R^I$ & $d^I$ & & $R^F$ & $d^F$ & & $\Delta$d \\
\midrule
        1MAG & 0.29 & & 6.5 & 4.95 & & 6.55\:(0.19) & 5.10\:(0.04) & & 0.14 \\ 
        1JNO & 0.20 & & 6.3 & 4.79 & & 6.33\:(0.07) & 4.87\:(0.06) & & 0.08 \\ 
Alanine tube & 0.21 & & 6.5 & 4.95 & & 6.52\:(0.05) & 4.99\:(0.04) & & 0.04 \\
\midrule
Structure (NSC) & RMSD & & $R^I$ & $d^I$ & & $R^F$ & $d^F$ & & $\Delta$d \\ 
\midrule
        1MAG & 0.28 & & 6.5 & 4.95 & & 6.57\:(0.05) & 5.11\:(0.04) & & 0.16 \\ 
        1JNO & 0.14 & & 6.3 & 4.79 & & 6.33\:(0.08) & 4.84\:(0.06) & & 0.05 \\ 
Alanine tube & 0.17 & & 6.5 & 4.95 & & 6.53\:(0.05) & 4.99\:(0.03) & & 0.04 \\
\bottomrule
\end{tabular}
\caption{RMSD between the initial (I) and optimised final (F) channnel geometries and corresponding 
helix properties (number of residues per helix turn $R$ and helix pitch $d$). Self consistent (upper)
and non selfconsistent (lower) results are shown.}   
\label{geometry}
\end{table}

\begin{figure}
\begin{center}
\includegraphics[width=0.5\textwidth]{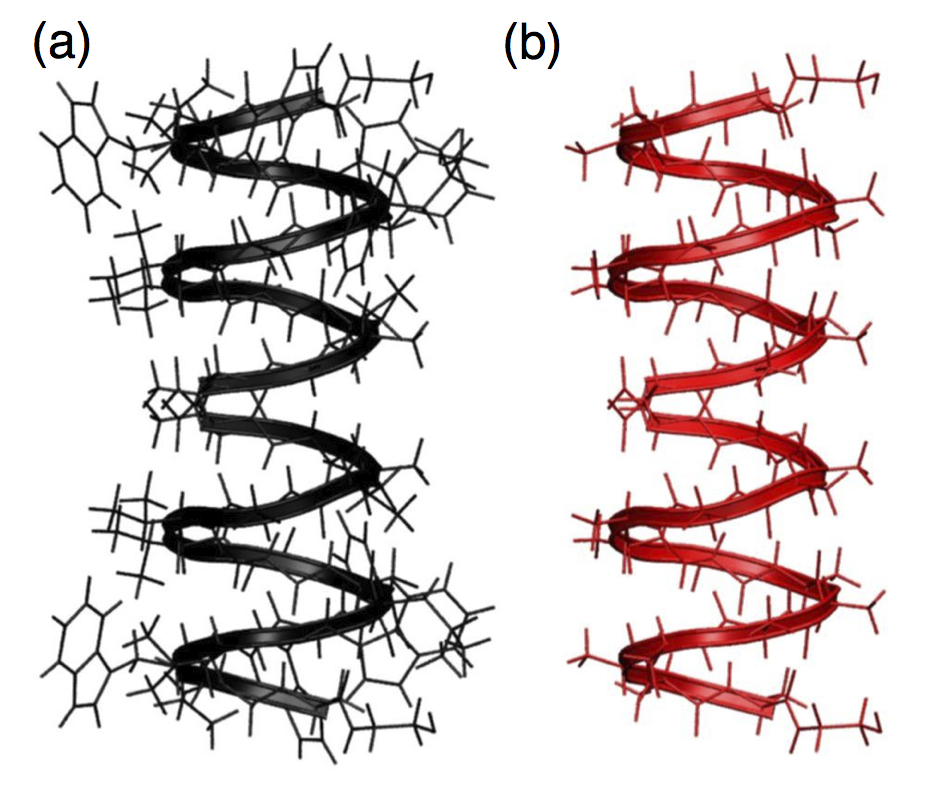}
\caption{ (a) Atomic structure of the gramicidin A (PDB code : 1MAG). (b) Alanine tube, which 
is made by replacing all the residues of gramicidin A by that of alanine.
(Online version in colour.)}
\label{ATgen}
\end{center}
\end{figure}

For the 1MAG structure, it is useful to know what causes the large change 
of the backbone structure during the optimisation: either the effect of backbone itself or 
that of side chains. 
One of the main differences between 1MAG and 1JNO structures is the different orientation 
of the indole ring of one Tryptophan (Trp9).
The atomic positions of other side chains are also slightly different.
These differences may cause the present different backbone structures. 
In order to investigate the effects of side chains on the backbone structure, 
we prepare a fictitious channel by replacing all amino acids of the gA with alanine,
keeping the same backbone structure as the original 1MAG structure,
shown in Fig.~\ref{ATgen}. 
Hereafter, we refer to this system as an alanine tube. 
We have relaxed the atomic positions of this system and analyse the values $R$ and $d$ of the 
resulting optimised structure.
As is shown in Table~\ref{geometry}, the change of the structure by the relaxation is found to be very small. 
The RMSD value is 0.21~\AA\ and the change of $d$ is only 0.04~\AA. 
This result suggests that the large change of the helix pitch during the structure 
optimisation of the 1MAG structure is due to 
the repulsion interactions between the side chains in the original 1MAG structure.

From the results so far, we can conclude that the original 1JNO structure is close to 
the optimised geometry in the gas phase, while 1MAG needs a large relaxation to reach 
the stable structure. 
By comparing the total energy, we find that initial 1JNO structure is more stable
than 1MAG by 10.3 eV. 
The corresponding energy difference calculated by classical force fields is
10.4 eV by CHARMM (v27) and 19.7 eV by AMBER (v99)\cite{note1}.
The value found using AMBER is larger than the DFT value, but these values are essentially 
consistent with the DFT result.
If the 1MAG structure is less stable than 1JNO by such a large amount of energy, 
it is unlikely that 1MAG structure will exist even in a lipid bilayer.
However, after the structure optimisation with DFT, this large energy difference between 
the two structures is greatly reduced.
The final structure of 1JNO is still more stable than 1MAG, but only by 1.8 eV. 
Although this energy difference is not negligible, it is small enough that the relative 
stability may change when we introduce the interactions between lipids and gA 
or include entropic effects, which are often important in soft biological molecules.
If the 1MAG structure gains more energy from these effects than 1JNO,
then the 1MAG structure might be more stable.
We note here that the thickness of the lipid bilayer is usually larger than the length of gA. 
Since the 1MAG structure is longer than 1JNO, the 1MAG structure may be more stabilised by the
interactions between gA and lipid molecules.  We will address this in future work.
However, we would like to emphasise here that the calculated structural properties of the 
isolated gA molecule in this work are useful in the analyses of such future works.

We now report the stabilisation energy of the two helices of gA.  
When the gA ion channel is open for ions, the upper and lower 
helices must be associated to form a dimer. 
They are also sometimes dissociated by the lateral movement in the membrane (lipid bilayers)
and in such cases the channel is closed for ion permeation.
Thus the stabilisation energy of the two helices can be regarded as a measure for 
the stability of the open channel structure.
In this respect, the stabilisation energy is important and we calculate the energy defined as
\begin{equation}
 E_{\rm s} = E_{\rm dimer} - 2E_{\rm mono}
\end{equation}
Here, $E_{\rm mono}$ is the total energy of the upper or lower peptides in isolation, 
whose atomic positions were relaxed.
$E_{\rm dimer}$ is the total energy of the associated gA, calculated with the basis set superposition 
error (BSSE) correction using the counterpoise method\cite{Boys:1970MP553}. 
With the optimised 1JNO structure, $E_{\rm s}$ is calculated to be 0.88 eV.
This value is reasonable given the number of hydrogen bonds between the two helices, 
which have three N-H---O and three O-H---N bonds. 
Note that the exchange-correlation functional of PBE has been 
reported to be accurate for the expression of hydrogen bonds\cite{Ireta:2004JPCA5692, 2008.otsuka.294201}.
In the membrane, this associated gA dimer structure will also be affected by 
the movement of surrounding lipid molecules. 
It will be important to compare this stabilisation energy with lipid-gA and 
lipid-lipid interactions in the future. 

Finally, we note that, while we have mainly discussed results obtained
with the SCF method, the differences between the NSC and SCF methods are small 
for the optimised structures (see Table~\ref{geometry}).
Since the NSC method is stable, and more efficient, it may play important
roles in the future study on larger and more complicated full gA systems.

\subsection{Electronic Structure}
\label{sec:electronic-structure}
In this section, we are particularly interested in the effects of the
side chains of the ion channel on the electronic structure 
in the pore region of the gA.
As mentioned in Sec.~\ref{sec:introduction}, it is known that the substitution of Trp significantly 
affects the rate of ion permeation\cite{Becker:1991BC8830,Hu:1995BC14147,Andersen_ja980182l, 2001.townsley.11676}.
It is important to understand whether such differences come from the different 
electronic structure of the side chains or some other factors, for example, from the structural
effects of the interactions of the side chains with lipid molecules. 
We note that there is a report showing that the conductance of a monovalent cation is
correlated with the electrostatic interaction energy between the ion and 
the dipole moment of the indole ring of the tryptophans\cite{Hu:1995BC14147}.

\begin{figure}[h]
  \centering
   \includegraphics[width=0.5\textwidth]{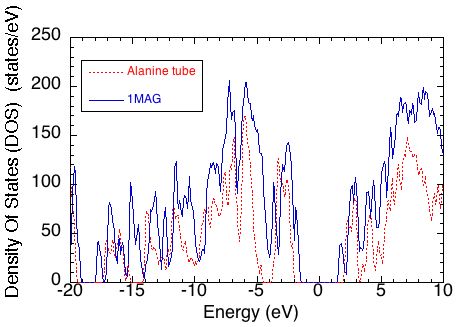}
  \caption{Electronic density of states for the alanine tube (red dotted line) and the 1MAG structure (blue solid line). 
  (Online version in colour.)}
  \label{fig:DOS}
\end{figure}

First, we show in Fig.~\ref{fig:DOS} the density of states of gA of 1MAG optimised
 structure (blue), as well as that of alanine tube (red) 
introduced in the last section.  We also calculated the DOS of the
1JNO structure, but the difference to the 1MAG DOS is small,
particularly around the band gap, so this is not shown for clarity.
We can see that the density of states of the 1MAG shows a HOMO-LUMO energy gap of 3.3 eV. 
The energy gap of alanine tube is much larger than that of gA, at around 4.2 eV, 
implying that HOMO and/or LUMO of gA mainly come from its side chains. 
The alanine tube also shows a very low DOS around 4eV below
the Fermi level, which is not present in the gA DOS.  

We studied the effects of side chains on the pore of 
the gA by investigating the charge density and the electrostatic potential. Ion transport depends critically on the electrostatic potential inside 
the channel pore, which in turn depends strongly on the charge density. 
The potential also affects the dynamics of the water molecules arranged in the pore.
If the charge density is plotted as a function of distance from 
the channel axis, i.e. the centre of the pore, the two systems are
found to be almost identical until the distance is greater than 
4.4~\AA, the approximate radius of the gA. 
We can thus conclude that the effects of the side chains on the charge distribution 
in the pore region are very small.

\begin{figure}
\begin{center}
\includegraphics[width=0.9\textwidth,clip=on]{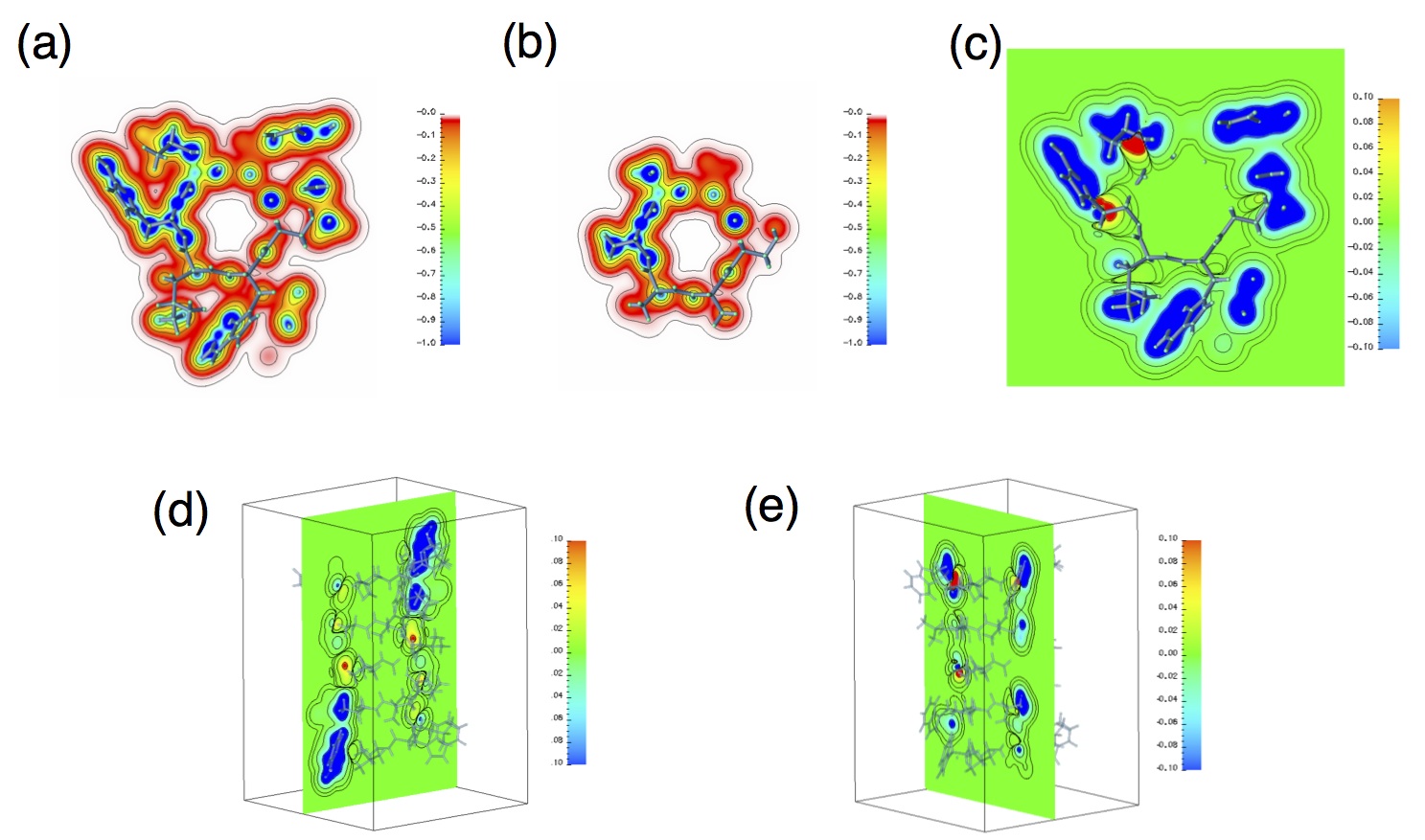}
\caption{DFT electrostatic potential [eV]: (a)-(c) are cross sections perpendicular to 
  the channel axis($z$) at 9.6 \AA\ from the centre of the channel.
  (a) 1MAG geometry; (b) alanine channel;(c) 1MAG and alanine channel potential
  difference.  Note the different scale in (c). 
  (d) and (e) are also potential difference between 1MAG and alanine channel, but 
  cross sections along the channel axis, $xz$ and $yz$ planes, respectively.
  The contours in (c)-(e) are drawn at 
  $10^{-1}, 10^{-2}, 10^{-3}, 10^{-4}, and 10^{-5}$ eV
  for both positive and negative values.
  (Colour figures.)} 
\label{potential}
\end{center}
\end{figure}

We next investigate the electrostatic potential, calculated as a sum
of the local pseudopotential 
and the Hartree potential using the self-consistent charge density.
Figure~\ref{potential}(a) shows a cross section of the potential perpendicular to the channel axis 
for the 1MAG structure,  9.6~\AA\  along the axis away from the centre of the gA. 
The binding site for the Tl$^+$(thallium) ion, which is similar to K$^{+}$ ion in size 
and hydration properties, is reported to be at this position\cite{1991.olah.847}.
The calculated binding site using 1MAG structure is also reported to be close to this 
position, 9.7~\AA\ \cite{Bastug:2006BJ3941}.
In the figure, areas of the lower potential are located near the peptide bonds which form hydrogen bonds.
In contrast, the region near the centre of the pore shows a high but flat potential, 
making it accessible to positively charged particles. 
Figure~\ref{potential}(b) shows the equivalent cross section for the alanine tube having a 1MAG backbone structure. 
There is a clear difference between Figs.~\ref{potential}(a) and (b) in the region of 
side chains, outside of the backbone.
However, the two potentials are almost the same in the pore region. 
Figure~\ref{potential}(c) shows the difference of these two potential surfaces, implying that the two 
pores exhibit identical electrostatic potential at least within 1 meV. 
The same behaviour can be seen in Figs. ~\ref{potential}(d) and (e),
which are cross sections in other planes, specifically the $xz$ and
$yz$ planes, showing the potential difference 
along the channel axis ($z$ axis).
We can conclude from these results that the effects of dipole moments of Trp groups are well screened and the
electronic structure of the side chains of the gA does not affect the electrostatic potential in its pore. 
 In order to explain the effect of Trp groups on the ion permeation rate, dynamical effects must be considered.
We suggest that the side chains of Trp groups may play important roles in the dynamics of the gA channel
through the interactions with water molecules and lipids, especially with their polar heads.
This aspect should be clarified in future work.

\subsection{Order-N calculations on the isolated gramicidin A}
\label{sec:order-n-calculations} 
We expect that the effects of the membrane or water molecules will be clarified by the all-atom
DFT simulations of the gA channel in a lipid membrane with bulk water
and counter-ions as the environment. 
We are now working to perform large-scale DFT simulations of the full gA system and 
the results will be reported in a future publication.
Once we can calculate the total energy and forces acting on atoms accurately with DFT 
for many snapshots of the full gA systems, and utilise the results with the so-called 
force matching method\cite{Ercolessi:1994dq}, 
it will be possible to generate a DFT-derived force field, with which we can investigate
dynamics of the biological system.
To perform such large-scale DFT calculations, we will need to use an $O(N)$ method.
As a preparation for these future studies, we have made preliminary $O(N)$ calculations on the 
isolated gA, in order to clarify the accuracy of the method on the gA
systems.  In particular, we can test the effect of the truncation of
the density matrix since we can perform exact calculations (we have
similarly tested convergence on a DNA ten-mer in
water\cite{2008.otsuka.294201} which found essentially complete
convergence for a range of 10\AA).
Here, we use the NSC method because the convergence behaviour will be qualitatively 
the same for the SCF methods. 
For the basis sets, contracted DZP basis sets\cite{2008.torralba.294206} are used.

\begin{figure}
\begin{center}
\includegraphics[width=0.7\textwidth,clip=on]{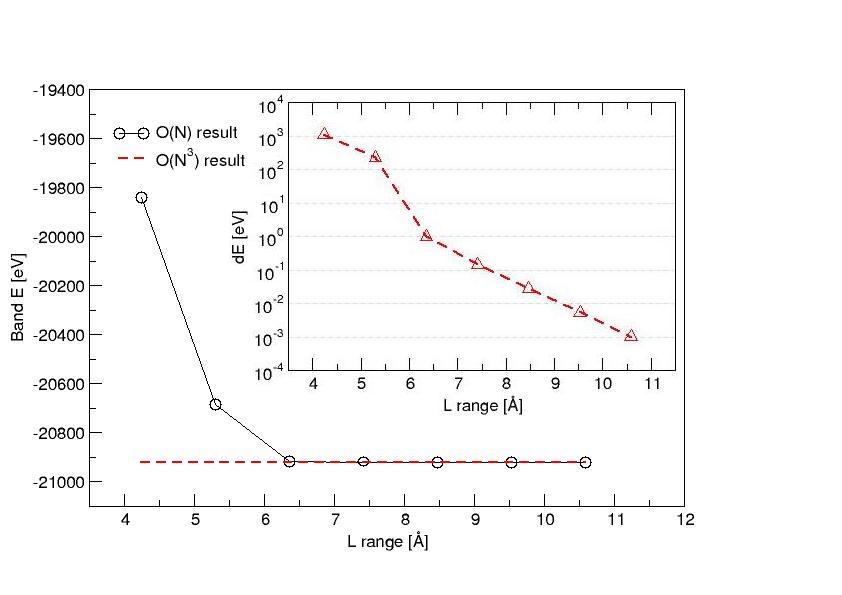}
\caption{Order-N band energy behaviour with increasing range of auxilliary density matrix $L$ (inset: convergence to diagonalisation result).
(Online version in colour.)}
\label{ondata}
\end{center}
\end{figure}

 Figure~\ref{ondata} shows the total energy calculated for different
 values of the cutoff, $R_L$. 
We find that all calculations are stable in the optimisation of $L$ matrix. 
Since the present system contains only 552 atoms, we can employ the exact diagonalisation method 
and the total energy calculated by the method is also plotted by a dotted line in the figure.
The inner panel of Fig.~\ref{ondata} shows the difference between the $O(N)$ and exact 
diagonalisation methods, on a log scale. 
This shows the total energy by the $O(N)$ method converges very fast towards the exact value. 
When the cutoff larger than 7.5~\AA\ is used, the energy discrepancy is less than 1\,meV per atom. 
This cutoff is small enough to realize actual calculations. 
In our previous O(N) DFT study of a different system, Ge nano structures on Si(001) surface, 
we actually used 10.8~\AA\ for the structure optimisation of the systems containing more than 20,000 atoms\cite{Miyazaki:2008tt}.

We expect that similar accuracy in the total energy would be obtained in the calculations 
of full gA systems, because the water and lipid molecules have large HOMO-LUMO energy gap,
resulting in fast convergence of the total energy with respect to $R_L$. 
In fact, in our investigations of dry and hydrated DNA systems, 
we found that the difference caused by bulk water is extremely small\cite{2008.otsuka.294201}. 
Based on these results, we expect that the $O(N)$ calculations of the full gA systems will be 
both highly accurate and useful for understanding structure and function.
These studies are already under way and we have recently succeeded in employing a self-consistent
DFT calculation on a full gA system, which consists of a gA channel embedded in 64 DMPC molecules,
and the water region made of 2473 water molecules ($\sim$ 16,000 atoms). 
The results will be presented in a future publication.

\section{Conclusions}
\label{sec:conclusions}

 In this work, we have performed a DFT study of gramicidin A (gA) to characterise the channel 
in isolation and to examine its electronic structure. 
In particular, we have investigated the effects of the side chains on
the structure and electronic properties of the channel.

For the structural stability of the gA in the gas phase, we have
performed DFT geometry 
optimisation starting from two different experimental atomic coordinates, 1MAG and 1JNO. 
We have found that the two calculations did not result in a common structure, but 
reinforced the difference in the helix properties, $R$ and $d$. 
Although the original 1MAG structure has larger helix pitch than 1JNO, 
the optimisation of 1MAG structure exhibited a helix expansion.
On the other hand, the optimised 1JNO structure in the gas phase was found to be very close to 
the original geometry.  
To investigate the effects of side chains on the helix structure, we have introduced an 
artificial alanine tube system, which has a common backbone structure with 1MAG but 
with all residues replaced by alanine.  
We have found that 
the optimised helix pitch of the alanine tube is close to that of the original 1MAG structure.
This result suggests that the original 1MAG structure includes
repulsion interactions between the side chains. 
Comparing the total energy of the optimised structures, the 1JNO structure is more stable 
than the 1MAG structure by 1.8 eV. 
However, this energy difference may become smaller in the membrane environment. 
Since the thickness of lipid bilayers is usually larger than the length of gA and
the 1MAG structure has smaller length mismatch, the 1MAG structure may be 
more stabilized than 1JNO by the interactions between gA and lipid molecules. 
This aspect is interesting and should be clarified in future studies. 
In addition to the structural optimization of gA, we have also calculated the 
stabilisation energy for the upper and lower helices to make a dimer. 
This energy is important because the gA system shows a gating behaviour for the 
ion permeation by association or dissociation of the two helices.
The energy is calculated to be 0.8 eV, implying that reasonably strong hydrogen bonds 
are formed between the two helices. 

 We have also investigated the electronic structures of gA and the alanine tube. 
The charge density and electrostatic potential in the pore of the two systems 
have been compared to clarify the effects of side chains. 
The potential is relevant for the ion transport and the dynamics of the water molecules 
arranged in a single file in the channel. 
We found that the potential in the pore is almost identical between the two systems. 
The results imply that the sensitivity of the channel function to the type of
residues cannot be explained by the electrostatic interactions made by
the dipole of Trp groups.
This does not support some of the proposals in previous reports. 

 In the next step, it is desirable to simulate the system in the channel environment, 
that is gA in membrane with bulk water. 
We have already started to work on this large and complex gA system containing 
more than 16,000 atoms using an O(N) DFT method.  
The code {\sc Conquest} used in this work has an ability to treat very large systems 
containing many thousands of atoms, with the density matrix minimisation method.
As a preparation for such large-scale DFT studies, we have applied the method to 
the isolated gA in this paper.
By calculating the total energy dependence on the cutoff of the auxiliary density matrix $R_L$,
we have found that the total energy converges very fast to the exact value. 
The errors from using the finite cutoff of $R_L$ become negligibly small if we use $R_L$ 
larger than 10~\AA. 
It is encouraging and we believe that the method will become one of the standard 
theoretical techniques to study the atomic and electronic structures of complex biological systems. 

\ack{The authors gratefully acknowledge useful discussions with
  Antonio Torralba, Michiaki Arita, Takao Otsuka and Wataru Shinoda.  
  Most of the calculations in this work were done using the numerical materials simulator at the National Institute 
  for Materials Science (NIMS), Japan.
  This work was partly supported by the Royal Society,  Grant-in-Aid for Scientific Research on Innovative Areas (No. 22104005),
  the Strategic Programs for Innovative Research (SPIRE), MEXT, and the Computational Materials Science
  Initiative (CMSI), Japan.
  }


\end{document}